\documentclass[prd,aps,superscriptaddress,floatfix,nofootinbib,eqsecnum,twocolumn]{revtex4-2}

\pdfoutput=1

\usepackage{amsmath}
\usepackage{amsfonts}
\usepackage{amsmath}
\usepackage{amssymb}
\usepackage{bm}
\usepackage{dcolumn}
\usepackage{graphicx}
\usepackage[latin1]{inputenc}
\usepackage{latexsym}
\usepackage{rotating}
\usepackage{hyperref}
\usepackage{subfigure}
\usepackage{color}
\usepackage{changes}
\usepackage{verbatim}
\usepackage{comment}
\usepackage{empheq}
\usepackage{csquotes}
\usepackage{physics}
\usepackage{float}
\usepackage{soul}
\usepackage{amsmath,latexsym}
\usepackage{mathrsfs}
\usepackage{orcidlink}

\usepackage{booktabs}  % For better horizontal rules

\begin{document}

%\title{A Relativistic Generalized Uncertainty Principle Framework for Field Theories}
\title{RGUP Corrections to Scalar and Fermionic Fields}
\author{Gaurav Bhandari\orcidlink{0009-0001-1797-2821}}\email{bhandarigaurav1408@gmail.com}\affiliation{Department of Physics,\\ Lovely Professional University, \\ Phagwara, Punjab, 144411, India}

\author{S. D. Pathak\orcidlink{0000-0001-9689-7577}}\email{shankar.23439@lpu.co.in}\affiliation{Department of Physics,\\ Lovely Professional University, \\ Phagwara, Punjab, 144411, India}

\author{Vikash Kumar Ojha\orcidlink{0000-0002-0641-4015}}\email{vko@phy.svnit.ac.in}\affiliation{Department of Physics \\ Sardar Vallabhbhai National Institute of Technology, \\ Surat, Gujrat,395007, India.}

\begin{abstract}
We investigate the Relativistic Generalized Uncertainty Principle (RGUP) effects on scalar and fermionic fields using the Stetsko-Tkachuk approximation. Modified equations of motion, Hamiltonians, and stress-energy tensors are derived in Minkowski spacetime, incorporating quantum gravitational corrections that ensure a minimal observable length and revert to standard dynamics when corrections are absent. For fermionic fields in curved spacetime, spin connections maintain gravitational consistency. This framework, applicable to high-energy physics, black hole thermodynamics, and cosmology, integrates quantum gravity into relativistic field theories.
\end{abstract}

%%%%%%%%%%%%%%%%%%%%%%%%%%%%%%%%%%
\maketitle

%\textbf{(Key words: Loop Quantum Cosmology beyond GR, Hamiltonian analysis, Brans-Dicke Theory, Bianchi spacetime)}

%%%%%%%%%%%%%%%%%%%%%%%%%%%%%%%%%%

%%%%%%%%%%%%%%%%%%%%%%%%%%%%%%%%%%

\section{Introduction}
The persistent challenge of formulating a complete theory of quantum gravity arises from the fundamental incompatibility between general relativity (GR) and quantum mechanics (QM). Despite their individual successes in describing macroscopic gravitational phenomena and microscopic quantum interactions, respectively, these frameworks remain mathematically irreconcilable \cite{c1,s1,h1}. Over the years, numerous approaches to quantum gravity have been proposed, each offering distinct advantages and limitations. Among the most prominent candidates are Loop Quantum Gravity (LQG) \cite{1,2,3,5}, string/M theory (ST) \cite{veneziano1986stringy,gross1988string,amati1989can,yoneya1989interpretation,s3}, and Doubly Special Relativity (DSR) \cite{6,7}. A common feature emerging from these theories is the presence of a fundamental minimal length scale, often associated with the Planck length. In string theory, this minimal scale corresponds to the finite size of a fundamental string, while in LQG, it manifests as a discrete area gap \cite{ashtekar2021short,rovelli2008loop,ra1}.

Given the ongoing development of a consistent quantum gravity framework, the notion of a minimum length has motivated various modifications to standard quantum mechanics. The existence of a minimal observable length leads to the generalization of the standard Heisenberg relation by incorporating gravitational-induced corrections, called the generalized uncertainty principle (GUP) \cite{17,18,19,20}. The GUP  extends the Heisenberg Uncertainty Principle (HUP) by incorporating quantum gravity corrections into the uncertainty relation. Inspired by results from LQG and string theory \cite{m1,si1}, the GUP introduces a momentum-dependent correction term that alters fundamental commutation relations. The GUP relation is written as

\begin{equation}
\Delta x \Delta p \geq \frac{\hbar}{2}\left[1+\beta_1 \left(\frac{l_p}{\hbar}\right)^2 (\Delta p)^2\right]
\end{equation}
where $l_p$ is the Planck length and $\beta_1$ is the positive numerical constant  \cite{e1,e2,bos1}. And, it is clear from the above relation that the minimal length $(\Delta X)_{min} = \sqrt{\beta_1} l_p $. 

In non-relativistic quantum mechanics, the HUP provides a bridge between classical and quantum descriptions by encoding the intrinsic measurement imprecision between position and momentum. Analogously, in the relativistic regime, the GUP is expected to play a similar foundational role, modifying quantum mechanical behavior at high energies. Beyond its conceptual significance, the GUP serves as an effective approach to incorporating quantum gravitational effects into physical models \cite{10,11,12,13,14,bh1,101,102,103,bh4,bat}. 

The predicted minimal length in quantum gravity is generally on the order of the Planck length ($l_{pl} = 10^{-35}$ m), where quantum gravitational effects are anticipated to become prominent. However, a significant challenge in integrating a minimal length into theoretical frameworks is its violation of Lorentz invariance, as the minimal length does not qualify as a Lorentz-invariant quantity. This constraint has limited the majority of GUP models to non-relativistic regimes, with only a few efforts made to extend GUP into relativistic contexts and Quantum Field Theory \cite{A1,A2,A3,A4,A5}. Recent research has investigated the generalized uncertainty principle (GUP) within a relativistic framework. Notably, \cite{B1} has introduced a relativistic generalized uncertainty principle (RGUP) that maintains Lorentz invariance while converging to the non-relativistic limit in low-energy scenarios. The establishment of this coherent relativistic GUP framework facilitates an examination of its implications for fundamental principles and elucidates how it modifies established results.

\section{Relativistic Generalized Uncertainty Principle (RGUP)}\label{RGUP}

The concept of a modified uncertainty principle was first introduced by Kempf, Mangano, and Mann in 1995 \cite{17}, where the standard position-momentum commutator is altered as follows:
\begin{equation}
[x^i, p^j] = i\hbar \delta^{ij} (1 + \beta_1 \mathbf{p}^2), \label{1}
\end{equation}
where $i, j \in \{1,2,3\}$. And using the Schr\"{o}dinger-Robertson uncertainty relation, one can easily obtain the minimal length uncertainty as $\Delta x_{\text{min}} = \hbar \sqrt{\beta_1}$. Extending this concept to a relativistic framework in Minkowski spacetime, a covariant generalization of the quadratic Generalized Uncertainty Principle (GUP), inspired by \cite{C1}, takes the following form
\begin{equation}
[x^{\mu}, p^{\nu}] = i \hbar (1 + (\epsilon - \alpha) \gamma^2 p^{\rho} p_{\rho}) \eta^{\mu \nu} + i \hbar (\alpha' + 2\epsilon) \gamma^2 p^{\mu} p^{\nu}, \label{2}
\end{equation}
where $\mu, \nu \in \{0,1,2,3\}$, and the metric signature is chosen as $(-,+,+,+)$. Here, $\gamma$ is a dimensionless parameter defined as $\gamma = \frac{1}{M_{\text{Pl}}c}$, where $M_{\text{Pl}}$ is the Planck mass. The parameters $\alpha$, $\epsilon$, and $\alpha'$ are also dimensionless and become significant at the Planck scale\cite{17,C1}. Also, one can observe that eqn.\eqref{2} naturally reduces to the non-relativistic form \eqref{1} in the limit $c \to \infty$ and recovers the standard Heisenberg uncertainty relation as $\gamma \to 0$. Since $x^{\mu}$ and $p^{\nu}$ are physical position and momentum for which one can introduce the canonical conjugate four vectors $x^\mu_0$ and $p^\nu_0$ as
\begin{equation}
p^{\nu}_0 = -i \hbar \frac{\partial}{\partial x_{0 \nu}}, \quad  [x^{\mu}_0, p^{\nu}_0] = i \hbar \eta^{\mu \nu}.
\end{equation}

Using results from \cite{B1,C1,g1}, we express the deformed position and momentum operators up to $\mathcal{O}(\gamma^2)$ as:
\begin{align}
  x^{\mu} &= x^{\mu}_0 - \alpha \gamma^2 p^{\rho}_0 p_{0 \rho} x^{\mu}_0 + \alpha' \gamma^2 p^{\mu}_0 p^{\rho}_0 x_{0 \rho} + \xi \hbar \gamma^2 p^{\mu}_0,\label{eqx}\\
    p^{\mu} &= p^{\mu}_0 (1 + \epsilon \gamma^2 p^{\rho}_0 p_{0 \rho}),
\end{align}
where $\xi$ is another dimensionless parameter. The commutator for the position operators then becomes
\begin{equation}
[x^{\mu}, x^{\nu}] = i \hbar \gamma^2 \frac{-2\alpha + \alpha'}{1 + (\epsilon - \alpha) \gamma^2 p^{\rho} p_{\rho}} (x^{\mu} p^{\nu} - x^{\nu} p^{\mu}).
\end{equation}

The last two terms in Eq.~\eqref{eqx} introduce directional dependence through $p^{\mu}_0$, which violates the principle of relativity. To preserve the principle of relativity, we set $\alpha' = \xi = 0$. The relativistic GUP then reduces to the KMM algebra in the $c \to \infty$ limit. For our purpose, we adopt the Stetsko-Tkachuk representation by choosing $\alpha = 0$, leading to the linearized modified algebra:
\begin{align}
[x^{\mu}, p^{\nu}] &= i \hbar (1 + \epsilon \gamma^2 p^{\rho} p_{\rho}) \eta^{\mu \nu} + 2 i \hbar \epsilon \gamma^2 p^{\mu} p^{\nu},\label{a1} \\
[x^{\mu}, x^{\nu}] &= 0.\label{a2}
\end{align}
From the above commutator relation, one can write the corresponding position and momentum operators as
\begin{align}
x^{\mu} &= x^{\mu}_0,\label{a3}\\
p^{\mu} &= p^{\mu}_0 (1 + \epsilon \gamma^2 p^{\rho}_0 p_{0 \rho}). \label{mp}
\end{align}
The physical squared four-momentum obeys the dispersion relation  in relativistic mechanics and quantum field theory is given as
\begin{equation}
p^{\rho} p_{\rho} = -(mc)^2,
\end{equation}
using eq.(\ref{mp}) and the above dispersion relation we obtain
\begin{equation}
p^{\rho}_0 p_{0 \rho} (1 + 2 \epsilon \gamma^2 p^{\sigma}_0 p_{0 \sigma}) = -(mc)^2,\label{3}
\end{equation}
Solving the above equation for $p^{\rho}_0 p_{0 \rho}$ and choosing only the solution which reduces to $-(mc)^2$ for the proper classical limit ($\gamma \to 0$), we find the modified dispersion relation
\begin{align}
p^{\rho}_0 p_{0 \rho} &= -\frac{1}{4 \epsilon \gamma^2} + \sqrt{\frac{1}{(4 \epsilon \gamma^2)^2} - \frac{(mc)^2}{2 \epsilon \gamma^2}}, \nonumber \\
&\simeq - (mc)^2 - 2 \epsilon \gamma^2 (mc)^4 + \mathcal{O}(\gamma^4). 
\label{rmo}
\end{align}
For this paper we restrict ourselves to this solution and use it for the relativistic GUP corrections to different classes of fields. one can also obtain the non-relativistic GUP by taking the limit ($c \to \infty$) such that the four-momentum scalar product reduces to
\begin{equation}
p^{\rho}_0 p_{0 \rho} = -\left(\frac{E}{c}\right)^2 + p^i_0 p_{i_0} \to - \hbar^2 \nabla^2. \label{b2}
\end{equation}
In the next section, we examine the application of the above equation to study the GUP corrections to the equation of motion, energy, and stress-energy tensor for different types of fields. From this point onward, we use $\epsilon \gamma^2= \beta$.

\section{RGUP corrections on different fields}

\subsection{Scalar field}

The dynamics of the field $\phi$ is governed by a Lagrangian  that depends on $\phi$, its time derivative $\dot \phi$, and the spatial gradient $\nabla \phi$. The Lagrangian density, which is the spatial integral of the Lagrangian, $L(t) = \int \mathcal{L}(\phi, \partial_\mu \phi)d^3x$, for a simple scalar field in Minkowski spacetime, is given as 
\begin{equation}\label{main1}
\mathcal{L}= \frac{1}{2}\partial^\mu \phi \partial _\mu \phi -V(\phi),
\end{equation} 
to study the dynamics in the presence of the RGUP, we incorporate its effects through a modified derivative operator. Specifically, we adopt a formulation where the position variable remains unchanged, while the momentum variable is modified as
\begin{align}\label{main2}
x^\mu \rightarrow X^\mu&=x^\mu ,\\ \nonumber
p^\nu \rightarrow P^\nu &= p^\nu(1-\beta (mc)^2)\equiv \partial ^\nu(1-\beta(mc)^2),
\end{align}
one can recover the non-relativistic version of the modified operator by taking the  $\lim c\rightarrow\infty$.
On substituting the modified operator from  Eq.(\ref{main2}) into scalar field Lagrangian, we obtain
\begin{equation}\label{mainG}
\mathcal{L'} = \frac{1}{2}\partial ^\mu \phi \partial_\mu \phi \{1-2\beta (mc)^2 + \mathcal{O}(\beta^2)\}-V(\phi),
\end{equation}
where the $\mathcal{L}'$ represents the RGUP modified Lagrangian density. One can obtain from the principle of least action, the Euler-Lagrange equation for the field $\phi$  as
\begin{equation}
\partial_\mu \frac{\delta \mathcal{L'}}{\delta(\partial_\mu \phi)}- \frac{\delta \mathcal{L'}}{\delta \phi}=0,
\end{equation}
which gives the equation of motion as
\begin{equation}\label{rgupkg}
(1 - 2\beta (mc)^2) \partial^\mu\phi \partial_\mu \phi+ V_{,\phi}(\phi) = 0,
\end{equation}
where the subscript $``,\phi "$ represent derivative wrt $\phi$.
The Eq.(\ref{rgupkg}) represents the RGUP-modified Klein-Gordon equation. The non-relativistic and standard KG equation counterparts can be recovered by taking the limits $c \rightarrow \infty$ and $\beta =0$, respectively. 

In a discrete system, one defines the conjugate momentum corresponding to each dynamical variable \( x \). When extending this formulation to a continuous system, the conjugate momentum is replaced by the conjugate momentum density \( \Pi \) associated with the field \( \phi(x) \), defined as:  
\begin{equation}
\Pi = \frac{\delta \mathcal{L'}}{\delta \dot \phi} = -\dot \phi (1-2\beta (mc)^2).
\end{equation}
From the Legendre transformation, the Hamiltonian density is defined as $H=\int [\Pi \dot \phi - \mathcal{L}]d^3x\equiv\int \mathcal{H} d^3x$, and thus one can obtain
\begin{equation}
\mathcal{H} = \Pi \dot \phi - \mathcal{L'} = -\frac{1}{2}(\dot \phi^2 + \Vec{\nabla} \phi)(1-2\beta (mc)^2)+V(\phi). 
\end{equation}

The \textit{stress-energy tensor} derived from the Lagrangian density as 
\begin{equation}
 T^{\mu \nu}  = \frac{\delta \mathcal{L'}}{\delta(\partial _\mu \phi)}\partial ^\nu \phi - \eta ^{\mu \nu} \mathcal{L'},  
\end{equation}
which gives
\begin{align}
T^{\mu \nu} &= \partial ^\mu \phi \partial^ \nu \phi (1-2 \beta(mc)^2) \nonumber \\
&\quad -\eta^{\mu \nu} \left[\frac{1}{2}\partial^\mu \phi \partial_\mu \phi(1-2\beta(mc)^2)-V(\phi)\right].
\end{align}

With the RGUP corrections, the stress-energy tensor also satisfies, 
\begin{widetext}
\begin{align}
    \partial_\mu T^\mu_\nu &= \partial_\mu \left[\partial ^\mu \phi \partial^ \nu \phi (1-2 \beta (mc)^2) 
    - \eta^{\mu \nu} \left(\frac{1}{2} \partial^\sigma \phi \partial_\sigma \phi (1-2\beta (mc)^2) - V(\phi) \right) \right] \notag \\
    &= \partial^2\phi \, \partial^\nu \phi (1-2\beta (mc)^2) 
    + \partial^\mu \phi \, \partial_\mu \partial^\nu \phi (1-2\beta (mc)^2)  
    - \partial^\nu \left[\frac{1}{2} \partial^\sigma \phi \partial_\sigma \phi (1-2\beta(mc)^2) - V(\phi) \right] \notag \\
    &= \left[\partial^\mu \phi \partial_\mu \phi (1-2\beta(mc)^2) + V'(\phi) \right] \partial^\nu \phi = 0, 
    \quad \text{(from the modified KG equation Eq.~\eqref{mainG})}.
\end{align}
\end{widetext}

The above results clearly show that after incorporating the RGUP corrections into the Lagrangian, the field retains all the standard properties of the original dynamics, and only the equation of motion for the scalar field acquires a correction term of the order of quantum gravitational parameter \( \beta \).

\subsection{Fermionic field}
In Minkowski spacetime, the Lagrangian density for a non-interacting spin-$\frac{1}{2}$ fermionic field $\psi$ is given by
\begin{equation}
\mathscr{L}=\tilde{\psi}(i \gamma^\mu\partial_\mu-M)\psi,
\end{equation}
$\psi$ and $\tilde{\psi}$ are the Dirac spinor field and its adjoint respectively, whereas $\gamma^\mu$ represents the Dirac gamma matrices.
Now including the RGUP corrections to the Lagrangian through modified momentum operator Eq.(\ref{main2}), we get
\begin{align}
    \mathscr{L}' &= \tilde{\psi} \left(i \gamma^\mu\partial_\mu(1-\beta (mc)^2)-M \right) \psi \\ \nonumber
    &= i\gamma^\mu\tilde{\psi}\partial_\mu\psi -\beta i \gamma^\mu  \tilde{\psi} \partial_\mu (mc)^2 \psi -M\tilde{\psi}\psi \\ \nonumber
    &= \mathcal{L}_{\text{Dirac}} -\beta i \gamma^\mu  \tilde{\psi} \partial_\mu (mc)^2 \psi.
\end{align}
And one can easily get the non-relativistic GUP result for Limit $c\rightarrow\infty$ as shown in the seminal work of \cite{nozari} as
\begin{equation}
\mathscr{L}' = i\gamma^\mu\tilde{\psi}\partial_\mu\psi -\beta i \gamma^\mu  \tilde{\psi} \partial_\mu (\grad^2) \psi -M\tilde{\psi}\psi.
\end{equation}
and without the quantum gravitational effects $\beta=0$ one can recover the original form of Lagrangian.

Using Euler-Lagrangian equations for the spinor field and its adjoint, we obtain
\begin{equation}\label{E1}
\frac{\delta \mathscr{L}'}{\delta \psi} = \partial_\mu  \frac{\delta \mathscr{L}'}{\delta (\partial_\mu \psi)},\quad \frac{\delta \mathscr{L}'}{\delta \tilde{\psi}} = \partial_\mu \frac{\delta \mathscr{L}'}{\delta (\partial_\mu \tilde{\psi})}
\end{equation}
From Eq.(\ref{E1}), we get
%\begin{equation}-\tilde{\psi} M = \partial_\mu (\tilde{\psi} i \gamma^\mu - \beta \tilde{\psi} i \gamma^\mu (mc)^2)\end{equation}or,
\begin{equation}
(\gamma^\mu\partial_\mu   -  M i- \beta \partial_\mu  (mc)^2)\tilde{\psi}  = 0.
\end{equation}
Similarly, for adjoint component 
\begin{equation}
(i\gamma^\mu \partial_\mu -M -\beta i \gamma^\mu \partial_\mu(mc)^2)\psi=0
\end{equation}

Now first defining the conjugate momenta
\begin{equation}
\Pi_\psi = \frac{\delta \mathscr{L}'}{\delta(\partial_0 \psi)} = \tilde{\psi} i \gamma^0 - \beta i \gamma^0 \tilde{\psi} (mc)^2,
\end{equation}
and
\begin{equation}
\Pi_{\bar{\psi}} = \frac{\delta \mathscr{L}'}{\delta(\partial_0 \bar{\psi})} = 0.
\end{equation}
The GUP-modified Hamiltonian obtained using Legendre transform gives
\begin{align}
\mathcal{H}' &= \Pi_\psi \partial_0\psi + \Pi_{\tilde{\psi}} \partial_0\tilde{\psi} - \mathscr{L}' \nonumber \\
&= \tilde{\psi} \left[-i\gamma^i\partial_i + \beta i \gamma^i \partial_i (mc)^2 + M \right] \psi. 
\end{align}
Since the above results are obtained for the Fermionic field in Minkowskian spacetime, in the next section, we extend our analysis by incorporating relativistic-GUP corrections in curved spacetime.

\subsubsection{Fermionic field in Covariant form }
In the curved spacetime, Lagrangian density for fermionic field \cite{book} is written as
\begin{equation}\label{curved}
\mathcal{L}=\frac{ei}{2}(\tilde{\psi}\gamma^\mu \mathcal{D}_\mu\psi - \mathcal{D}_\mu \tilde{\psi}\gamma^\mu \psi)-eM\tilde{\psi}\psi,
\end{equation}
where $e\equiv \det(e^\mu_a)$,the determinant of the vierbein, which relates curved-space volume elements to flat-space. The curved spacetime couterpart of $\gamma$ matrices is $\gamma^\mu=e^\mu_\alpha\gamma^\alpha$ and also satisfies,
\begin{equation}
\{\gamma^\mu, \gamma^\nu\}=2g^{\mu\nu}.
\end{equation}
In curved spacetime, the ordinary derivatives are replaced with covariant derivatives, which incorporate spin connections to maintain consistency with the underlying geometry. The covariant derivative can be read as 
\begin{align}
    \mathcal{D}_\mu \psi &= \partial_\mu \psi + \Gamma_\mu \psi, \\
    \Gamma_\mu &= -\frac{1}{4} \omega_{\mu}^{ab} \gamma_a \gamma_b,
\end{align}  
where \( \Gamma_\mu \) represents the spin connection, and \( \omega_{\mu}^{ab} \) denotes the spin connection coefficients associated with the local Lorentz frame. Now incorporating the RGUP-modified covarient derivative as
\begin{equation}
\mathcal{D}_\mu \equiv \mathcal{D}_\mu(1-\beta (mc)^2),
\end{equation}
and substituting into the Lagrangian density Eq.(\ref{curved}) gives 
\begin{widetext}
\begin{align}
\mathcal{L}' &= \frac{ei}{2} \left( \tilde{\psi} \gamma^\mu \mathcal{D}_\mu (1-\beta (mc)^2) \psi - \mathcal{D}_\mu (1-\beta (mc)^2) \tilde{\psi} \gamma^\mu \psi \right) 
- eM \tilde{\psi} \psi \nonumber \\
&= \frac{ei}{2} \left( \tilde{\psi} \gamma^\mu \mathcal{D}_\mu \psi - \mathcal{D}_\mu \tilde{\psi} \gamma^\mu \psi \right) 
- \frac{\beta (mc)^2 ei}{2} \left( \tilde{\psi} \gamma^\mu \mathcal{D}_\mu \psi - \mathcal{D}_\mu \tilde{\psi} \gamma^\mu \psi \right) 
- eM \tilde{\psi} \psi.
\end{align}
\end{widetext}
Similarly, from the Euler-Lagrangian equation:

\begin{equation}\label{EG1}
\frac{\delta \mathcal{L}'}{\delta \psi} = \mathcal{D}_\mu  \frac{\delta \mathcal{L}'}{\delta (\mathcal{D}_\mu \psi)}, \quad \frac{\delta \mathcal{L}}{\delta \tilde{\psi}} = \mathcal{D}_\mu \frac{\delta \mathcal{L}'}{\delta (\mathcal{D}_\mu \tilde{\psi})} .
\end{equation}
we get
%\begin{equation}
%\begin{aligned}
%-\frac{ei}{2} \mathcal{D}_\mu \tilde{\psi} \gamma^\mu + \frac{\beta (mc)^2 ei}{2} \mathcal{D}_\mu \tilde{\psi} \gamma^\mu - eM\tilde{\psi} &= \mathcal{D}_\mu \left(\frac{ei}{2} \tilde{\psi} \gamma^\mu - \frac{\beta (mc)^2 ei}{2} \tilde{\psi} \gamma^\mu \right), \\
%\mathcal{D}_\mu \tilde{\psi} \gamma^\mu - \beta (mc)^2 \mathcal{D}_\mu \tilde{\psi} \gamma^\mu - iM\tilde{\psi} &= 0,\\ 
%-\gamma^\mu \mathcal{D}_\mu \psi + \beta (mc)^2 \gamma^\mu \mathcal{D}_\mu \psi -Mi\psi&=0.\end{aligned}\end{equation}
\begin{align}
\mathcal{D}_\mu \tilde{\psi} \gamma^\mu - \beta (mc)^2 \mathcal{D}_\mu \tilde{\psi} \gamma^\mu - iM\tilde{\psi} &= 0, \\ 
-\gamma^\mu \mathcal{D}_\mu \psi + \beta (mc)^2 \gamma^\mu \mathcal{D}_\mu \psi -Mi\psi&=0. 
\end{align}
where  we use $\mathcal{D}_\mu e = e \Gamma^\lambda_{\lambda \mu}$ and $\mathcal{D}_\mu \gamma^\mu = - \Gamma^\lambda_{\lambda\mu } \gamma^\mu$.
To obtain the Hamiltonian density, we define the conjugate momentum associated with the field as   
\begin{align}
    \Pi_\psi &= \frac{\partial \mathcal{L}'}{\partial(\mathcal{D}_0\psi)} = \frac{ei}{2} \tilde{\psi} \gamma^0 (1-\beta(mc)^2), \\
    \Pi_{\tilde{\psi}} &= \frac{\partial \mathcal{L}'}{\partial(\mathcal{D}_0 \tilde{\psi})} = \frac{ei}{2} \gamma^0 (1-\beta(mc)^2) \psi,
\end{align}
which gives the RGUP-modified Hamiltonian density upto the first order $\beta$ in curved spacetime as
\begin{align}
\mathcal{H} &= \Pi_\psi \mathcal{D}_0\psi + \Pi_{\tilde{\psi}} \mathcal{D}_0\tilde{\psi} - \mathcal{L}' \nonumber \\
&= \frac{ei (1-\beta (mc)^2)}{2} \left[\tilde{\psi} \gamma^i \mathcal{D}_i \psi + \mathcal{D}_i \tilde{\psi} \gamma^i \psi \right] + e M \tilde{\psi} \psi.
\end{align}

One can also derive the RGUP-deformed stress-energy tensor as
\begin{equation}
T^\mu_\nu = \frac{\partial \mathcal{L}'}{\partial(D_\mu \psi)}D_\nu \psi+\frac{\partial \mathcal{L}'}{\partial(D_\mu \tilde{\psi})}D_\nu \tilde{\psi}-g^\mu_\nu \mathcal{L}'
\end{equation}
which gives 
\begin{equation}
\begin{aligned}
    T^\mu _\nu &= \frac{i}{2} \tilde{\psi} \gamma^\mu (1-\beta (mc)^2) \mathcal{D}_\nu \psi 
    - \frac{i}{2} \gamma^\mu \psi (1-\beta (mc)^2) \mathcal{D}_\nu \tilde{\psi} \\
    &- g^\mu_\nu \bigg\{ \frac{i}{2} \big( \tilde{\psi} \gamma^\alpha \mathcal{D}_\alpha (1-\beta (mc)^2) \psi  
    - \mathcal{D}_\alpha (1-\beta (mc)^2) \tilde{\psi} \gamma^\alpha \psi \big) \\
    & - M \tilde{\psi} \psi \bigg\}.
\end{aligned}
\end{equation}
It is observed that the RGUP introduces higher-order $\beta$ terms in Hamiltonian density, which eventually will modify the energy spectrum and quantization of fermions in curved spacetime.
\section{Conclusion} \label{con}
The effects of quantum gravity are inherently universal, influencing the dynamics of diverse fields and potentially guiding the emergence of new physics at high-energy scales. Since quantum gravitational corrections become significant in extreme energy regimes where classical gravity and quantum mechanics intersect, it is crucial to investigate their impact on field dynamics. Among various phenomenological approaches, the GUP provides a compelling framework to incorporate minimal length corrections, offering insights into quantum gravity-induced modifications to fundamental interactions.
In this work, we have systematically examined the relativistic GUP-induced corrections in the dynamics of both scalar and fermionic fields. The interplay between relativistic and quantum mechanical phenomena at high-energy scales underscores the necessity of a comprehensive framework that integrates these effects. The RGUP provides such a framework, allowing for a consistent incorporation of quantum gravity-induced modifications alongside relativistic corrections. We incorporate these modifications into the field equations, and observe that how the deformation parameters affect equations of motion, energy-momentum structure, and conservation laws of these fields. Our analysis reveals that GUP-induced corrections alter the standard equations of motion and behavior of the fields.  Furthermore, we observe that RGUP corrections provide a more generalized and complete framework for understanding dynamics under quantum gravity effects. This formulation naturally encompasses the non-relativistic GUP limit by taking $c \to \infty$ and recovers the standard, unmodified dynamics in the absence of minimal length effects by setting the GUP parameter $\beta = 0$. This study not only extends previous investigations on GUP-modified scalar and fermionic fields but also provides a foundation for exploring further implications in quantum cosmology, black hole physics, and high-energy phenomenology. Future research could involve incorporating these corrections into interacting field theories, gauge field dynamics, and exploring their potential observational signatures in astrophysical and cosmological settings.

%In this paper, we systematically investigate the RGUP-induced corrections in classical field theory. We employ the Stetsko-Tkachuk approximation to obtain the deformed position and momentum operators and apply them to both scalar and fermionic fields. Specifically, we modify the Klein-Gordon and Dirac Equations, analyze the Effects on Field Dynamics and extend RGUP modifications to fermionic field in Curved Spacetime.

%%%%%%%%%%%%%%%%%%%%%%%%%%%%%%%%%%%%%%%%%%%%%%%%%%%

%%%%%%%%%%%%%%%%%%%%%%%%%%%%%%%%%%%%%%%%%%%%%%%%%%%

\end{document}